\documentclass[jmp,amsmath,amssymb,reprint]{revtex4-1}
\usepackage{graphicx}
\usepackage{dcolumn}
\usepackage{bm}

\begin{document}

\title{Bloch-Siegert shift in application to the astrophysical determination of the fundamental constants variation}

\author{D. Solovyev}
\email[]{solovyev.d@gmail.com}
\affiliation{V. A. Fock Institute of Physics, St. Petersburg
State University, Petrodvorets, Oulianovskaya 1, 198504,
St. Petersburg, Russia}

\date{\today}

\begin{abstract}
We have evaluated the Bloch-Siegert shift for the different values of magnetic field's strengths defined at astrophysical conditions, i.e. when the stars with the strong surface magnetic fields are taken as a powerful pumping source of radiation. It is found that the additional shift of resonant frequency should be taken into account in the search for the time variation of the fundamental constants. The main conclusion is that the influence of the electromagnetic field shift should be considered carefully in each special case of the corresponding frequency determination.
\end{abstract}

\maketitle

{\it Introduction.} - The search for a potential time variation of fundamental constants has been the focus of theoretical and experimental investigations since this idea was first suggested by Dirac \cite{Dirac},\cite{Dirac2}. Various limits or evidence reported from astrophysics, cosmology and laboratory experiments, as well as evidence from the natural nuclear fission reactor in
Oklo (Africa) are still under discussion. Recently, renewed interest to the problem has been triggered by the first positive results concerning the variation of the fine-structure constant $\alpha$, obtained from quasar spectra \cite{Webb}. The status of the problem was reviewed in \cite{Uzan} (see also \cite{Alb}) and corresponds to the previous decade. Nevertheless, the reported results should be interpreted with care \cite{Bah}. In this respect it is important to stress that during recent years the accuracy of the laboratory experiments has demonstrated especially fast progress. It was reported that the variation of $\alpha$ expressed as the
logarithmic derivative was of the order of $\frac{\dot{\alpha}}{\alpha} \approx 10^{-15}$ yr$^{-1}$ \cite{Webb}, while it is claimed that in the newly proposed experiment utilizing the radio-frequency transition in atomic dysprosium \cite{Ng} an accuracy of the order of $\frac{\dot{\alpha}}{\alpha} \approx 10^{-18}$ yr$^{-1}$ can be achieved, with the dot signifying a derivative with respect to time.

Initial attempts to measure the variation in $\alpha$ were based on the method (AD method) which uses the difference in the wavelengths of the doublets originating from the same ground state (i.e., ${}^2 S_{1/2} \rightarrow {}^2 P_{3/2}$ and ${}^2 S_{1/2}\rightarrow {}^2 P_{1/2}$ transitions). The constraints on the variation in $\alpha$, $\Delta\alpha/\alpha$, are obtained by assuming that the measured difference in the wave-length centroid of the doublets is proportional to $\alpha^2$ to the lowest order. The best constraint obtained using this method is $\Delta\alpha/\alpha = (-0.5 \pm 1.3) \times 10^{-5}$ \cite{Chand}. Like absorption doublets one can also use the central wavelength of multiple emission lines originating from the same initial excited state. For example, authors \cite{Bah} use the nebular OIII emission lines at $\lambda$5007 and $\lambda$4959 originating from ${}^1D_2$ excited level and derive $\Delta\alpha/\alpha = (0.7 \pm 1.4) \times 10^{-4}$ based on 73 quasar spectra over the redshift range $0.16 \leq z \leq 0.80$. Studies based on molecular absorption lines seen in the radio/mm wavelength range are more sensitive than those based on optical/UV absorption lines. The results of a detailed many-multiplet analysis of a new sample of Mg II systems observed in high quality quasar spectra obtained using the Very Large Telescope were presented in \cite{Sria}. The variation in $\alpha$ over redshift range $0.4 \leq z \leq 2.3$ derived in \cite{Sria} is $\Delta\alpha/\alpha = (-0.06 \pm  0.06) \times 10^{-5}$ and $-2.5\times 10^{-16}\leq (\Delta\alpha/\alpha\Delta t)\leq +1.2\times 10^{-16}$ yr$^{-1}$.

On the other hand the laboratory experiments have placed stringent, model-free constraints on $\alpha$ variation (see \cite{Kozlov} and references therein): $\dot{\alpha}/\alpha=(-1.6\pm 2.3)\times 10^{-17}$ yr$^{-1}$. The crucial question for the proposed experiment \cite{Ng} is the possibility of storing the light during the essential time period (about 1 month). The most straightforward way to achieve a storage would be the use of a system of mirrors. Therefore, the question arises under which conditions multiple reflections (scatterings) can lead to the frequency shift, comparable with a relative shift of the order $10^{-15}$ \cite{LSPAS}. In 2008, Rosenband et al. \cite{Ros} used the frequency ratio of Al$^+$ and Hg$^+$ single-ion optical atomic clocks to place a very stringent constraint on the present time variation of $\alpha$, namely $\Delta\alpha/\alpha = (−1.6\pm 2.3)\times 10^{-17}$ per year.

One of the methods of the $\alpha$ variation detection consists in the investigation of the quasars spectra and comparison of the wave-length of the distinctive spectral lines with the corresponding modern values \cite{Webb2}. Quasars are used as a universal source of light in the most distant universe. Interstellar gas clouds in galaxies absorb the light emitted by these quasars. Thus in the resulting spectra the absorption line arises that can be attributed to known elements. Using the Keck telescope and a data set of 128 quasars at redshifts $0.5 < z < 3$, Webb et al. found that their spectra were consistent with a slight increase in $\alpha$ over the last 10 - 12 billion years. They found that
\begin{eqnarray}
\label{1}
\frac{\Delta\alpha}{\alpha}\equiv \frac{\alpha_{prev}-\alpha_{now}}{\alpha_{now}}=(-5.7\pm 1.0)\times 10^{-6}.
\end{eqnarray}
 
Quasar absorption spectra at 21 cm and UV wavelengths were used to estimate the time variation of $x\equiv \alpha^2 g_p\mu$, where $g_p$ is the proton $g$ factor, and $m_e/m_p\equiv \mu$ the electron/proton mass ratio \cite{Tz}-\cite{CFK}. Authors \cite{Tz} searched the optical data for heavy-element absorption features close to the redshifts where there is 21 cm absorption. 

In this paper we investigate the influence of the electromagnetic field on the absorption line frequency, namely, the Bloch-Siegert shift \cite{Bloch} is evaluated. The atomic/molecular system considered as a two-level system constructed from two hyperfine sublevels. The Bloch-Siegert shift is a phenomenon in quantum physics that becomes important for driven two-level systems when the driving gets strong (e.g. atoms, driven by a strong laser drive).

{\it Bloch-Siegert shift.} - The derivation of the Bloch-Siegert shift can be found, for example, in \cite{Stenh}. In accordance to \cite{Wei}, \cite{Wei2}, we can write (in SI units)
\begin{eqnarray}
\label{2}
\frac{d}{dt}\rho_{11}=i\chi \cos\omega t (\rho_{21}-\rho_{12})
\nonumber
\\
\frac{d}{dt}\rho_{22}=-i\chi \cos\omega t (\rho_{21}-\rho_{12})
\\
\nonumber
\frac{d}{dt}\rho_{21}=-(i\omega_{21}-\gamma_{21})\rho_{21}-i\chi\cos\omega t(\rho_{22}-\rho_{11})\, ,
\end{eqnarray}
where $V_{21}=V_{12}^*=-\mu_{12}B\cos\omega t=-\hbar\chi\cos\omega t$, $\mu_{12}$ is the magnetic dipole moment of the transition $|2>\rightarrow |1>$, $\hbar$ is the Planck constant, $B$ is the magnetic field, $\chi$ is the Rabi frequency, $\omega_{12}=(E_2-E_1)/\hbar$, $\gamma_{21}$ represents the one half of the level width $\Gamma_2$ in our case.

Adopting Stenholm’s formalism \cite{Stenh} to derive a solution (we omit all the details of the corresponding derivation due to their wide use), one can find that the maximum of the lorentzian line profile will be shifted and the transition frequency will be defined by (in atomic units)
\begin{eqnarray}
\label{3}
\omega=\omega_{21}+\frac{(\mu B)^2_{12}}{4\omega_{21}}
\end{eqnarray}
A similar result can be received for the multi-photon transition \cite{Stenh}. The external field $B$ can be considered in both cases: classical and quantum.

The matrix element $(\mu B)_{12}$ can be estimated as $(\mu B)_{12}\sim \mu_B B$, where $\mu_B=\frac{e\hbar}{2m_e}$ is the Bohr magneton, and $e$, $m_e$ are the electron charge and mass. Thus the energy, including the Bloch-Siegert shift can be written in the form
\begin{eqnarray}
\label{4}
\hbar\omega=\hbar\omega_{21}\left(1+\frac{(\mu_B B)^2}{4(\hbar\omega_{21})^2}\right).
\end{eqnarray}

The hydrogen 21 centimeter wavelength emission occurs from a rare quantum phenomena known most commonly as spin-flip transition. Spin-Flip transition occurs from a transition of a higher energy state to a lower energy state of an $n = 1$ neutral hydrogen atom with the frequency $\nu=1420405751.800\, Hz$. 

{\it Magnetic field definition.} - We define the magnitude of the magnetic field $B$, assuming that the absorption line for the hydrogen atom is observed in the direction of the radiation source (quasar). The dipole magnetic field waves propagate outwards, taking with them some energy. This energy is taken from the rotational energy of the neutron star. The magnetic field's strength on the star surface reaches $B_*\approx 3\times 10^{8}$ $T$. For different estimates \cite{Shkl}-\cite{Kant} the value of the emission power is about $10^{31}-10^{38}$ $erg/s$. Wave radiation more intense the greater the angle $\phi$ between the axis of rotation and the magnetic axis, with $\phi = 0$, i.e. parallel to these axes, no radiation. For the fast rotating Crab pulsar, PSR 0531 + 21, which is the quazi-ortogonal rotator ($\phi = 87^\circ$) with large magnetic field (period $P = 33$ ms, $B_* = 7\times 10^{12}$ $G$), the emission power was evaluated as $10^{35}$ $erg/s$ \cite{Kant} that also does agree with observations \cite{Mal}.

Energy carried away by the waves per time, or the power of the magnetic dipole radiation, is given by \cite{Shkl}, \cite{Ost}
\begin{eqnarray}
\label{5}
W_B=\frac{2 B^2_*}{3c^3\mu_0}\Omega^4R^6\sin^2\phi.
\end{eqnarray}
Here $c$ is speed of light, $\Omega$, $R$ are the angular frequency and the radius of the neutron star, $\mu_0$ is the vacuum permeability. 

The intensity of the field radiation on the distance $r$ from the source can be defined from
\begin{eqnarray}
\label{6}
I=\frac{W_B}{4\pi r^2}.
\end{eqnarray}
From the other hand, we can write
\begin{eqnarray}
\label{7}
I=\frac{cB^2}{8\pi\mu_0}\, ,
\end{eqnarray}
where $B=B(r)$ is the field strength at the distance $r$.

The inverse-square law generally applies when energy is radiated outward radially in three-dimensional space from a point source. The angular size of the source PSR 0531 + 21 can be easily calculated via $\varphi=2\arctan 2R/D\sim 10^{-12}$ $rad$ ($R$ is the star radius $R\approx 12.5km$ and $D$ is the nebula diameter $D\sim 3.4$ $pc$), i.e. the radio-emission source PSR 0531 + 21 can be considered as a point one within the Crab nebula (the angular size of the source less then $2''$).                                                              

Expression Eq. (\ref{6}) (inverse-square behavior) reflects, in principle, the dilution coefficient 
\begin{eqnarray}
\label{8}
\mathit{w} = \frac{I}{I_*} = \frac{1}{2}\left(1-\sqrt{1-\left(\frac{R}{r}\right)^2}\right)\, ,
\end{eqnarray}
where $I_*$ is the field intensity on the surface of the star.

For the $r\gg R$
\begin{eqnarray}
\label{9}
\mathit{w} \approx \frac{1}{4}\left(\frac{R}{r}\right)^2\, ,\, {\rm i.e.}
\end{eqnarray}
and from Eq. (\ref{7}) it follows
\begin{eqnarray}
\label{10}
 B=\frac{1}{2}\frac{R}{r}B_*.
\end{eqnarray}

{\it Some examples and discussion.} - Consider, as example, the Crab nebula. Different estimates give the value of the emission power about $10^{31}-10^{38}$ $erg/s$. To evaluate the distance between the source and the absorber we use $r\approx D/2\sim 1.7 pc$, which leads us to the $I\approx 2.89\times 10^{-7}$ $W/m^2$ for the $10^{35}$ $erg/s$ and, therefore, $B\approx1.74\times10^{-10}$ $T$, Eqs. (\ref{6})-(\ref{7}). This field gives a small value of Bloch-Siegert shift of the order of $7.39\times 10^{-19}$. For the case if $W_B\sim 10^{38}$ we obtain $B\sim 5.52\times 10^{-9}$ $T$ and the Bloch-Siegert shift of the order $7.4\times 10^{-16}$, Eq. (\ref{4}). On the other hand, the radio sources associated with the old remnants of supernovae, have a shell structure, i.e. they have radio-emitting region located on the periphery. Nothing like this in the distribution of Crab Nebula was observed. Radio sources in this case, fill the entire volume of the nebula, concentrating toward the edge. In fact at the distance $r\sim (10^{-5} - 10^{-4})D$ we have the field's strength on an absorber of the order of $10^{-4} - 10^{-3}$ $T$ and $\mu_B^2 B^2/4(h \nu)^2\sim  10^{-7} - 10^{-5}$ that is the same order of magnitude as the expected variation of fine structure constant $\alpha$. 

In an environments ionized by radiation with a spectrum close to the Planck, e.g. in planetary nebula, areas of $HII$, and for a number of elements ($C$, $Si$, $S$, $Fe$, etc.), in regions of the interstellar medium of $HI$, the ionization equilibrium of the substance depends on radiation dilution. In this case, the ionization is much lower than that is via the Saha formula, since the right-hand side of the Saha formula should be multiplied by $\mathit{w}$. Typical values ​​of $\mathit{w}$ in the specified regions $\sim 10^{-12}-10^{-16}$. Thus for the magnetic field strength on absorber we obtain the magnitude $B$ of the order $\sim 10^{-4}-10^{-8}$ $T$, with the estimate $B_*\sim 10^8$ $T$, Eq. (\ref{9}). Then we can conclude that the Bloch-Siegert shift lies in interval $10^{-7} - 10^{-15}$.

An important feature of the compact light source is its variability, particularly strong in the case of emission of $H_2O$. In a few weeks or even days of the line profiles quite different. Sometimes significant variations occur in 5 minutes, which is possible only if the dimensions of the sources does not exceed the distance that light travels in that time (otherwise statistical fluctuations will be reimbursed). Thus, the size of the regions emitting $H_2O$ lines may be about 1 astronomical unit ($au$). On the distance from the source about 1 $au$ we obtain $B\sim 0.00193541$ $T$ and, therefore, the Bloch-Siegert shift of the order $9.1\times 10^{-5}$ for the hydrogen atom.

Observations show that in the same area with dimensions of a few tenths of a parsec could be many sources, some of which only emits $OH$ lines, and some - only lines $H_2O$. Only known in physics until the emission mechanism that can give tremendous power within only a narrow range of the spectrum, is coherent (i.e. the same phase and direction) light lasers, which are called optical lasers, and radio - masers. Compact sources of emission of $OH$ and $H_2O$ are likely gigantic natural cosmic masers. The typical size of the maser clusters is about $10^{14} - 10^{15}$ $m$ and the neutron star radius is of the order of $10 km$. Thus, the radiation dilution coefficient is equaled approximately $2.5\times 10^{-23} - 2.5\times 10^{-21}$  and, therefore, $\mu_B^2 B^2/4(h \nu)^2\sim 2.4\times 10^{-5} - 2.4 \times 10^{-7}$ for the hydrogen line $21$ $cm$ and $1.7 \times 10^{-5} - 1.7 \times 10^{-7}$ for the $OH$ $18$ $cm$ line.

On the other hand we can employ the data reported in \cite{Will}. We use the values of the radiation fluxes $f_{OH}$, which is related to depth of spectral line profile, for the field's strength definition. Defining the redshift we can evaluate the distance to the source $\frac{c z}{H_0}$ (for the $z\lesssim 1$ and Habble constant $H_0\approx 2.28\times 10^{-18}$ $s^{-1}$) and, therefore, to estimate the possible Bloch-Siegert shift for the $HO$ molecule at the $\lambda=18$ $cm$. In order to approximate the distance between the source and absorber we use the cluster size $10^{14} - 10^{15}$ $m$. The results of calculations are presented in Table 1.
\begin{table}
\caption{\, Table 1. In first column the names of objects are given. Second column contains the values of the radiation flux in $10^{-21}$\,\,$\frac{W}{cm^2}$, third one denotes the redshifts of the source. Fourth column corresponds to the values of the magnetic field strength obtained via Eq. (\ref{7}). The values of $B_{abs}$ in $10^{-6}T$ are listed for the distances $10^{14}-10^{15}$ $m$, respectively. The last column represents the values of the Bloch-Siegert shift $\frac{(\mu_B B)^2}{4(h\nu)^2}$ multiplied by the $10^{-9}$ factor for the corresponding distances.}
\begin{center}
\begin{tabular}{c|c|c|c|c}
\hline
\hline
Name $OH$M & $f_{OH}\times 10^{-21}$   & $z_*$ & $B_{abs}$, $\mu T$& BS shift \\
 IRAS & \,\,$W/cm^2$  & \qquad & \qquad & $\times 10^{-9}$\\
 \hline
01418+1651 & -6.1 & $0.0274$ & $25.7-2.57$ & $11.7-0.117$ \\
\hline
13428+5608 & -28 & 0.0378 & $76-7.6$ & $102-1.02$ \\
\hline
15327+2340 & -172 & 0.0181 & $90.4-9.04$ & $144-1.44$ \\
  \hline
  \hline
\end{tabular}
\end{center}
\end{table}

{\it Conclusions.} - Bloch-Siegert shift was considered in this paper in view of the search of the fundamental constant variation. The atomic system (hydrogen atom or $OH$ molecule) was considered as a two-level system constructed from two hyperfine sublevels subjecting to the magnetic field incident from a powerful radio-source. Such models have been of interest over the years for applications to physical systems, including spin problems and atoms in strong electromagnetic fields. The expression of the Bloch-Siegert shift can be found, for example, in \cite{Nov}, where the BS shift was observed experimentally in a system of optically oriented $C^{133}$ atoms. More approximate expression is presented in this paper, Eq. (\ref{4}), and can be easily found in literature.

In astrophysics quasars, neutron stars or radio-masers (megamasers) can serve a source of the strong magnetic fields. The magnitude of the magnetic field's strength on the star surface reaches $B\approx 3\times 10^{8}$ $T$ and energy carried away by the waves at a time, or the power of the magnetic dipole radiation of star, can be found Eq. (\ref{5}). On different estimates the value of the emission power is about $10^{31}-10^{38}$ $erg/s$. Thus such stars can be considered as powerful pumping sources.

It is found that the shift of resonant frequency can be significant from point of view of modern investigations in the search of fundamental constant variation. The magnitude of the field's strength can be found via Eqs. (\ref{6}), (\ref{7}) or radiation dilution coefficient Eq. (\ref{9}). In principle, the simple estimates can be done: $1a.u.\approx 4\times 10^8$ $G$ and the line 21 cm corresponds to the frequency $2.17\times 10^{-7}$ in atomic units. For the field about 1 $G$ we have $\mu_B B\sim 0.25\times 10^{-8}$ $a.u.$ Thus the Bloch-Siegert shift, Eq. (\ref{4}), at such fields reachs the value about $3.32\times 10^{-5}$ or is the same order as Eq. (\ref{1}).

The most accurate definition of frequencies can serve for the estimates of the time variation of the fundamental constants via the comparison of the present and past values Eq. (\ref{1}). A number of tiny effects should be taken into account for this purpose. On our mind the Bloch-Siegert shift can be one of them. Some estimates of the Bloch-Siegert shift at the astrophysical conditions can be found through the paper (see Table 1 also). The definition of the distance between the source and absorber is quite uncertain, but can operate for the rough evaluation. The main conclusion of this paper is that the detailed analysis of the magnetic field influence is required for the each special case of the frequency determination.

\begin{center}
Acknowledgments
\end{center}
The work was supported by RFBR (grants No. 08-02-00026,  No. 11-02-00168-a and No. 12-02-31010) and goskontrakt No. 8420. The author is grateful to Prof. L. N. Labzowsky and Prof. V. K. Dubrovich for valuable discussions.

\end{document}